\documentstyle[12pt]{article}
\newcommand{\cent}{\centerline}
\newcommand{\vs}{\vspace*}

\begin{document}

\baselineskip 0.6cm

\begin{center}

{\large {\bf Ettore Majorana's Inaugural Lecture on Theoretical Physics}}

{ (Neaples, Italy; Jan.13, 1938)}

\

{\large {\em with comments by the editors, Bruno Preziosi and
Erasmo Recami:}} \ \protect\footnote{E-mail addresses for
contacts: recami@mi.infn.it [ER]; preziosi@unina.it [BP]}

\end{center}

\cent{ Bruno Preziosi, }

\vs{1mm}

\centerline{{\em Universit\`a di Napoli, Neaples, Italy;}}
 \cent{{\rm and} {\em INFM-Sezione di Napoli, Neaples, Italy.}}

\vs{1mm}

\centerline{\rm and}

\vs{1mm}

\cent{ Erasmo Recami }

\vs{1mm}

\cent{{\em Universit\`a di Bergamo, Bergamo, Italy;}}
\cent{{\rm and} {\em INFN---Sezione di Milano, Milan, Italy.}}

\vs{6mm}

{{\bf Abstract of the English version \ --} \ Ettore Majorana was
a member of Enrico Fermi's research group in Rome, Italy. Fermi
did regard Majorana as much brihter than himself as far as
theoretical physics was concerned (more information can be found
particularly in the arXives' e-print physics/9810023, in Italian,
and refs. therein, and also in the recent multilanguage
arXiv:0708.2855v1 [physics.hist-ph]). In 1937 Majorana
partecipated in the national Italian competition, for a chair in
theoretical phyics, requested by Emilio Segre' at that time at
Palermo University: Other competitors being GC.Wick, G.Racah, and
G.Gentile jr.  After a proposal of the judging Commette, chaired
by E.Fermi, Majorana got a full-professorship at Naples
University, for exceptional scientific merits, outside the
competition normal procedures. In this e-print we make known the
notes prepared by Majorana for his Inaugural Lecture (and
discovered long ago, in 1973, by one of the present authors (ER)),
together with some comments: {\bf Both in English} (first article)
{\bf and in Italian} (second article, with a short Bibliography at
its end). The present articles have been prepared on the occasion
of the Centenary (2006) of Majorana's birth. The present
preliminary notes for the Inaugural Lecture reveal Majorana's
interest not only for scientific research, but also for the best
didactical methods to be followed in order to teach classical and
quantum physics in the most effective way (while his approach to
Special Relativity is known to us from his lecture notes,
published elesewhere). \ Let us seize the present opportunity also
for recalling that almost all the biographical documents regarding
Ettore Majorana (photos included) are protected by copyright in
favour of Maria Majorana together with one of the present editors
[ER] (and with the publisher Di Renzo [www.direnzo.it]), and
cannot be further reproduced without the written permission of the 
right holders.}\\

\newpage

{\bf ETTORE MAJORANA:}

{\large{\bf Preliminary notes for the inaugural lecture}}

\large{\rm University of Naples, 13 January 1938}

\

In this first introductory lecture I will briefly discuss the
aims of modern physics and the significance of its methods, with
particular emphasis on their most unexpected and original aspects with
respect to classical physics.

Atomic physics, which will be the main subject of my discussion,
despite its important and numerous practical applications ---together with
those of a wider and perhaps revolutionary impact that the future may
have in store---, is first of all a science of immense
\textit{speculative} interest
for the depth of its investigation that really reaches the extreme roots of
natural facts. Let me first mention, without referring to any
specific category of experimental facts and without the help
of mathematical formalism, the general characters of the conceptions
of nature that the new physics has introduced.

\begin{center}
* * *
\end{center}

As is well known, at the beginning of our century the \textit{classical
physics} of Galileo and Newton is entirely founded on a \textit{mechanistic}
conception of nature that from physics has spread out not
only to the sciences that are closer to it but also to biology
and even to social sciences. This conception was extended in very
recent times to almost all the scientific thinking and to a good extent
also to the philosophical one, even
though, to tell the truth, the usefulness of the mathematical
method, which represented the only valid justification of the
mechanistic conception, has always been limited only to physics.

This conception of nature rested essentially on two pillars:
the objective and independent existence of matter, and physical
determinism. As we shall see, in both cases these notions were based
on common experience and were then generalized and given a universal
and infallible character mostly because of the irresistible fascination
that the exact laws of physics have always had even on the deepest
thinkers: they were considered as a sign of the absolute and
the revelation of the essence of the universe whose secrets,
as Galileo already proclaimed, are written in mathematical characters.

The \textit{objectivity} of matter derives, as I have said, from common
experience which teaches us that material objects have their
own existence independently of the fact that they are or are
not observed. Classical mathematical physics has added to this elementary
observation the further statement or requirement that it
is possible to give a mental representation of this objective
world which is perfectly adequate to explain reality; and that such a
mental representation can consist in the knowledge of a series
of numerical quantities sufficient to determine
at every point in space and at every instant
of time the state of the physical universe.

\textit{Determinism} instead only partially derives from common experience.
In fact this common experience gives contradictory indications:
besides facts that inevitably occur, as for example
the free fall of a body in vacuum, there are others ---and not only in the biological
world--- for which the inevitable occurrence is at least little evident.
Determinism, as a universal principle of science, could
therefore be formulated only as a generalization of the laws of
celestial
mechanics. It is well known that a \textit{system} of points
---as the bodies of our planetary system can be considered because
of their enormous distances--- moves and changes according to Newton's law.
This law states that the acceleration of one of these points is
obtained from the sum of as many vectors as the other points are:\looseness-1

$$
\ddot{\overrightarrow{P_r}} \propto \Sigma_s\frac{m_s}{%
R_{rs}^2}\overrightarrow{e_{rs}},
$$

$m_{s}$ being the mass of a generic point and $\overrightarrow{e}_{rs}$
the unit vector with direction from $\overrightarrow{P}_{r}$
to $\overrightarrow{P}_{s}.$
If we have a total of $n$ points, $3n$ coordinates will be necessary
to fix their position, and Newton's law establishes among these
quantities as many second-order differential equations whose
general integrals contain $6n$ arbitrary constants.
These constants can be determined by assigning the position and the
velocity components of each point at the initial time. Hence
it follows that the future configuration of
the \textit{system} can be predicted by calculation, provided we know
its initial state, \textit{i.e.} the set of positions and velocities of
the points which compose it. Everyone knows the extreme accuracy
with which astronomical observations have confirmed the exactness
of Newton's law and how astronomers can actually predict with its help
only, and
even in the distant future, the precise instant of an eclipse
or a conjunction of planets or other celestial events.

\begin{center}
* * *
\end{center}

To illustrate the present state of \textit{quantum mechanics} there exist
two almost opposite methods. One is the so-called historical
method: It explains how, starting from precise and almost immediate
experimental indications,
the first idea of the new formalism was born; and how its
subsequent development was compulsorily determined more by its internal
consistency than by newly discovered fundamental experimental phenomena.
The other method is the mathematical one, according to which
the quantum formalism
is presented right from the beginning in its most general and
therefore clearest structure and only later its criteria of
application are discussed. Each one of these two methods, if exclusively applied,
has very serious drawbacks.

It is a fact that, when quantum mechanics was born, for some
time is was looked at by many physicists with surprise, skepticism
and even considered as completely incomprehensible. This was mainly
due to the fact that its logical consistency,
internal coherence and sufficiency appeared dubious and even elusive.
This was also attributed, though in a completely wrong way, to a special obscurity of exposition
by the first founders of the new mechanics. But the truth is that
they were physicists and not mathematicians and for them the evidence
and justification of the theory rested essentially on the immediate
applicability to the experimental facts that had suggested it.
The general formulation, clear and rigorous, came later partly
thanks to mathematical minds. If we were then to simply repeat the exposure
of the theory according to its historical appearance,
we would unnecessarily create at first an uncomfortable or distrustful feeling
that was justified in the old days but which can no longer be
accepted and can be spared.
Furthermore physicists ---who have managed to clarify,
not without trouble, the quantum methods by means of conceptual
experiments imposed by their own historical progress--- have, not rarely,
felt the need for a greater logical coordination
and a more perfect formulation of the principles, and have not refused
the help of mathematicians in this effort.

The second method, the purely mathematical one, presents even greater
inconveniences. It does in no way allow to understand the origin of the formalism
and as a consequence the place that quantum mechanics has in the history
of science. Moreover it does not fulfil at all the desire
to somehow perceive by intuition its physical significance, often so easily
satisfied by classical theories; finally its applications, though
quite numerous, appear few and disconnected, and even modest
compared to its overwhelming and incomprehensible generality.

The only way to make life easier for those who begin today the
study of atomic physics, without any sacrifice of the historical origin
of the ideas and even of the language we use today, is to start
with an ample and clear discussion of the mathematical tools
that are essential to quantum mechanics. Then the student will
be already familiar with such tools, when the time will come
to use them, and will no longer be frightened or surprised
by their novelty: at this stage, one will be thus able to proceed rapidly
to derive the theory from the experimental data.

Most of these mathematical tools already existed before the beginning of
the new mechanics (they had been without specific interests introduced by
mathematicians who
did not forecast such an exceptionally wide field of application);
but quantum mechanics has ``forced'' and extended them
to satisfy its practical needs. Thus we will expose them not as mathematicians
but rather as physicists would do, \textit{i.e.} with the criterion
not to worry about an excessive formal
rigour, which is not always easy and often totally
impossible to achieve.

Our only ambition will be to discuss as clearly as possible the
effective way in which physicists have been using those tools for
over a decade: It is this use, that has never led to any difficulty
or ambiguity, that constitutes the essential source of their certainty.

\newpage


\qquad
\vskip50pt

\cent{\Large{\bf COMMENTS ON THE ``PRELIMINARY NOTES}}

\cent{\Large{\bf FOR THE INAUGURAL LECTURE.''}}

\vskip40pt
\setcounter{footnote}{0}

\section*{\textbf{1. Majorana: the appointment to the chair and his \textit{lectio
magistralis}}}
\vspace{-5pt}

After the 1926 competition, in which Fermi, Persico and Pontremoli
were appointed professors, ten years passed before a new
competition for theoretical physics was announced in 1937; it was
required by the University of Palermo on the initiative of Emilio
Segr\'{e}. Ettore Majorana too decided to apply to the competition
(either on his own initiative or because he was invited to do so
by his friends). This decision may appear strange to those who
know Majorana's temper, who was so far away from academic
interests. However, we got an explanation during the last few
months. Let us first recall that when Majorana came back from
Leipzig at the end of 1933, he took the distances from Fermi's
group, but not from physics, as testified[4] by a large number of
documents\footnote{{E. Recami}, \textit{Il Caso Majorana:
Epistolario, Documenti,
    Testimonianze} (Mondadori, Milan) 1987, 1991;
see the IV updated edition (Di Renzo Editore, Rome) 2002.}
Moreover[5], De Gregorio\footnote{{A. De Gregorio} and {S.
Esposito}, in \textit{Sapere}, no. 3, June 2006, 56; see also
\textit{Teaching theoretical physics: The cases of E. Fermi and
  E. Majorana}, preprint \texttt{arXiv:physics/0602146}.}
has recently discovered, at the University of
Rome ``La Sapienza'', that during the time when he lived in isolation,
namely in the
academic years 1933/34, 1934/35 and 1935/36, Majorana had asked
for the opportunity 
to deliver a ``free'' university course at
the Institute in via Panisperna, which was his right since
he was ``Libero Docente'', namely qualified for university teaching. His requests were approved
by the Director Corbino, but Majorana never gave any lecture,
probably because of the lack of students able, at that time, to
understand the importance of his lectures.

Majorana was very interested in teaching what his prodigious mind
was understanding and discovering about the laws of nature. It is
likely that he applied to the competition just to finally have his
own students (whom he always took care of, as we are going to
see). As we know, after the proposal of the commission in charge
of the competition, chaired by Fermi, on November 2nd 1937, the
Minister Bottai issued the act of appointment for Majorana as full
professor of theoretical physics, at the Royal University of
Naples, out of competition. At the end of 1937, Majorana was
informed by the Minister of this appointment at his dwelling place
in viale Regina Margherita 37, in Rome, with this explanation:
\textit{``for the high reputation You achieved in the study of the
mentioned discipline, as from November 16th 1937-XVI}.'' Majorana
went to Naples after the Epiphany (around January 10th 1938), and
on January 12th he writes from his university seat to Minister
Bottai[6] saying, among other things, \textit{``I wish to affirm
that I shall devote all my energies to the Italian school and
science, today in such a successful ascension}\footnote{The
documents, discovered and published for the first time by E.
Recami are present in ref.[4]. They all are protected by copyright
(including the photographs) held by E. Recami, the Majorana family
and presently by the publisher Di Renzo. Any reproduction of these
documents is forbidden without written permission of the copyright
holders (exception is obviously made for the \textit{scientific}
papers).}.''\looseness-1\vspace{-0.7pt}

Majorana's letters, which are relevant also for the circumstances
in which Majorana disappeared, are contained in ref.[4]. We will
briefly mention only those which are of interest in this section.
In his letter to his mother dated January 11th 1938 from Naples,
Majorana wrote: \textit{``I announced the beginning of the course
for next Thursday 13th at nine. But it hasn't been possible to
check if there is overlapping with other classes, thus it is
likely that students will not come and we will need to postpone
the beginning of the course. I spoke to the dean and we agreed on
avoiding any formal character to the inauguration of the course,
also for this reason I should suggest you not to come\ldots''}.
Majorana's family, on the contrary, was punctual on Thursday
January 13th 1938, at nine, at Majorana's inaugural lecture. The
professors of the faculty were present at the lesson, among whom
certainly Antonio Carrelli and Renato Caccioppoli, who were very
good friends; as Gilda Senatore recalls, students were not
invited.\vspace{-0.7pt}

The notes for the opening address to the course, or inaugural
lecture, have been discovered by one of us [E.R.] about 1972 and
have been made public for the first time[7,8] ten years
later\footnote{{E. Recami}, in \textit{Corriere della Sera}
(Milano), 19 Ottobre 1982. See also refs.[4] and [8].}
\footnote{B. Preziosi (Editor), \textit{Ettore Majorana - Lezioni
all'Universit\`{a} di Napoli}
 (Bibliopolis, Napoli) 1987. The volume contains, besides a comment by
N. Cabibbo, also an article by E. Recami which includes the notes
for the inaugural lecture and a catalogue of all the unpublished
scientific papers by Majorana (edited by M. Baldo, R. Magnani and
E. Recami); for the catalogue see also {E. Recami}, in
\textit{Quaderni di Storia della Fisica}, no. \textbf{5} (1999),
19-68, e-print physics/9810023}.In the mentioned notes all the
interest of the scientist appears, not only for the general and
basic issues which animate the scientific research, but also for
the best \textit{didactic method} to follow in order to pass on
his knowledge to the students (for whom he had a deep care).

The notes of Majorana's inaugural lecture reveal several aspects
of his scientific and human qualities. We wish to warn that they
concern classical physics and quantum mechanics: in this first
stage relativistic aspects were neglected. These aspects are
examined in the second half of Majorana's course, as revealed by
the notes of his last six lessons which have been recently
discovered. Majorana was so particularly fascinated by the
antimechanicistic and probabilistic description of quantum
mechanics, that he discusses[9] it widely also in his posthomous
article\footnote{{E. Majorana}, ``Il valore delle leggi
statistiche nella fisica e nelle scienze sociali'',
\textit{Scientia} \textbf{36} (1942) 58-66. }, published in 1942
by Giovannino Gentile. On July 27th 1934 from Monteporzio Catone
(Rome) he had written to Gentile himself: \textit{``I think that
the major merit of} (10] \textit{Jeans' book}\footnote{{J. Jeans},
\textit{I Nuovi Orizzonti della Scienza} (Sansoni, Firenze) 1934,
Italian translation by G. Gentile jr. } \textit{is to anticipate
the psychological reactions which the new development of physics
will fatally produce when everybody understands that science has
stopped being a justification for the vulgar materialism.''}

Since the myth has associated the death of Majorana with
the fear of the construction of the atomic bomb, we can say immediately
that from the very beginning of his inaugural lecture Ettore says
openly: 
\textit{``Atomic physics, which will be the main subject of my discussion,
despite its important and numerous practical applications ---together with
those of a wider and maybe revolutionary impact that the future may
have in store---, is first of all a science of immense
\textit{speculative} interest
for the depth of its investigation that really reaches the extreme roots of
natural facts.''}
Majorana's words
make us understand that, despite the probable ``revolutionary''
applications which nuclear and atomic physics could have led to,
they interested him mainly from the speculative point of view.

\vspace{-3pt}

\section*{\textbf{2. Lectures in his theoretical physics course}}
\vspace{-3pt}

As testified by Gilda Senatore and Sebastiano Sciuti, his course
was attended, from January 15th onwards, by themselves and by Nella
Altieri, Laura Mercogliano, Nada Minghetti and Savino Coronato,
the latter being one of Caccioppoli's students; after the last
lecture he no longer attended the Physics Institute and took
his degree in mathematics in the same year. Nobody else attended
his lectures, apart from a few occasional presences of Mario
Cutolo, who had already graduated in physics.

Majorana took much care of his students, and appreciated them. In
fact, in the last letter to his friend and colleague Giovanni
Gentile jr, he wrote: \textit{``I am satisfied with my students;
some of them seem determined in approaching physics seriously}.''
When the chalk was in his hands, his bashfulness disappeared and
entire blackboards were easily filled with elegant physical and
mathematical symbols. This behaviour has been recalled by Gilda
Senatore in a TV interview performed by Bruno Russo, and, more
recently, at the meeting organized by the University of Naples
Federico II on the occasion of the 60th anniversary of his
disappearance.

All the autograph notes of his lectures, written with care by
Majorana, as aids for his students (maybe Ettore had in mind to
write a book for his students, in the same way as he did when he
wrote his original study notes[11], the
\textit{``Volumetti''}\footnote{{S. Esposito},
  {E. Majorana} jr, {A. van der Merwe} and {Recami E.},
\textit{Ettore Majorana - Notes on Theoretical Physics} (Kluwer
Academic Press, Dordrecht, Boston e New York) 2003. (Edition in
the original Italian language: {E. Majorana}, {\textit{Appunti
inediti di fisica teorica},} edited by Esposito S. e Recami E.
(Zanichelli, Bologna) 2006.).}), were left in safekeeping,
the day before going to Palermo, to his favorite pupil Gilda
Senatore, together with some other writings which were no more
found. A letter by Preziosi published in ``Le Scienze'' (September
2002), and[12] ref.~\footnote{\textit{L'eredit\`{a} di Fermi e
Majorana ed altri temi} (Bibliopolis, Napoli) 2006.} explain how
those lecture notes happened to be in Carrelli's hands between the
end of 1938 and the beginning of 1939, and in which occasion were
then sent to E. Amaldi, unfortunately without six lessons on
electrodynamics and special relativity. It is interesting to
notice that in $1939-\!1940$ Carrelli taught special relativity
and the related notes were published by GUF in 1940. The notes
sent to Amaldi and deposited at the Domus Galilaeana were
published anastatically in ref.[8].

Recently[13], S. Esposito\footnote{S. Esposito, \textit{Nuovo
Saggiatore}, \textbf{21} No. 1-2 (2005) 21-41.} and A. Drago have
found, among the papers left by Eugenio Moreno, a student in
Mathematics who took his degree with Caccioppoli in 1941, a
personally handwritten copy of all Majorana's manuscripts,
including the part relative to special relativity, absent in the
papers deposited at the Domus Galilaeana. Such notes[14], now, are
all present in ref.~\footnote{S. Esposito (Editor), \textit{Ettore
Majorana - Lezioni di Fisica Teorica} (Bibliopolis, Napoli)
2006.}.

\vspace{-6pt}

\section*{\textbf{3. The procedures for the chair assignment
and for the \textit{lectio magistralis} in Neaples}} \vspace{-3pt}

On June 5th 1224, Fredric II, king of Germany and emperor of the
Romans, sent forth from Siracuse to all authorities in the Kingdom
a circular letter (\textit{generales licterae}) which commenced:

\textit{``With God's
blessing, for whom we live and reign, to whom we report all the
good we do, we wish that, in our Kingdom, through a source of
science and a breeding ground of erudition, many may become wise and
shrewd, who, made skilful by the meditation and the study of
law, may serve God, to whom all things serve, and be useful to us
for the worship of justice, whose commands we order all of you
to obey. Therefore we have provided that, in the most pleasant city
of Naples, arts may be taught and studies of all professions
may be cultivated, thus those who are thirsty and greedy for
erudition may find within the Kingdom itself how to satisfy their
thirst, not being forced to procure education, by embarking on long
journeys and begging in foreign lands.''}

In the same circular he announced that \textit{``one of the
scholars he meant to choose was the well learned Roffredo di
Benevento''} and stated that \textit{``loans will be allowed to
pupils\ldots''}; as referred[1] by Torraca\footnote{Stamperia di
Giovanni de Simone, Napoli MDCCLIV.}, lectures at the ``Studium''
began in October 1224.\looseness-1

The appointment of the professors was then a privilege of the
Emperor. This procedure was kept by Corrado and Manfredi and,
after a brief interruption following the battle of Benevento (1226),
by the Angevins (1266-1442) and later by the Aragoneses (1442-1503).
In 1503, with Ferdinand the Catholic, the Spanish period started.
At the beginning the Studium was closed for a few years, but
it was reopened on St. Luca's day (October 18) in 1507. It is interesting
to point out that the University of Salamanca, funded six years
earlier than the Studium, used to open on the same day; moreover,
as we shall see further down, the University of Salamanca was
always taken as a reference institution.\looseness-1

The person who brought important innovation was certainly the
viceroy D. Pietro Fernandez de Castro, Earl of Lemos. Not only
did he order that a big building was built outside Costantinopoli
gate ---the building hosted the Studium from 1615 till the beginning
of the XVIII century and at present it hosts the archaeological
\mbox{museum---,} but he also made a deep reform in the way professors
were recruited between 1614 and 1616. The reform is basically
the same as the rule sanctioned by the University of Salamanca
in 1561, which states that the assignment is carried out through
a competition announced by the government and after a public
examination in front of a commission made of professors and lecturers
even belonging to religious orders, but, contrary to Salamanca's procedure,
there were no students. In Salamanca, after the public examination,
the competitors were invited to wait in a chapel for the call of
the winner and the invitation to join the professor board.

As Giangiuseppe Origlia reports[2] in his \textit{Istoria dello
Studio di Napoli}\footnote{F. Torraca, \textit{Storia della
Universit\`{a} di Napoli} (Riccardo Ricciardi Editore, Napoli)
MCMXXIV.}, in the public exam the applicant was \textit{``imposed
to explain publicly and loudly for the duration of one whole hour
without the help of any written paper\ldots\ those topics of the
subject\ldots\ which had been given to him 24 hours earlier by the
Prefect in the presence of witnesses.''}

To underline the interest with which these lessons were attended,
Origlia says that the people who attended the lessons were \textit{``lecturers
and all those who had the right to vote for the chair which was
to be assigned, as well as a crowd of students, and others,
who wished to attend such `jousts'.''}

Earl of Lemos also fixed the rules for the opening of the academic year
(the first time on June 14th 1615). According to the testimony
of a person of the time, the ceremony started with a procession \textit{``opened
by the jurists, wearing a green brocade
`mozzetta'{\renewcommand{\thefootnote}{*}\footnote{Short cloack.}} and a hat
with a green flock; the physicians, with a blue brocade `mozzetta'
and a hat with a blue flock were in the middle; at the end the
theologians with a white brocade `mozzetta' and a hat with a
white flock}.''
\setcounter{footnote}{2}

Once the procession reached the palace of the Studium, the ceremony
started with a \textit{lectio magistralis} read by Gio. Lorenzo Rogiero.

Although the clothes drew sniggers from some people, they
were used also in the following analogous ceremonies.

In ref.[2], Nino Cortese describes the occasions in which an
official lecture was read: \textit{``The academic year commenced
solemnly with an oration of one of the lecturers; moreover, a
custom which was common in the previous century was saved
according to which when one of the lecturers was appointed to a
chair he had to read a real opening lecture.}''

The opening lecture was a custom common in Salamanca too and
it became a tradition which was formalised in Spain with a Royal
Decree on 20th August 1859, whereas in Naples it remained an internal
rule of the University.

Going back to the assignment procedure, we must say that it
did not always take place through a public competition; in fact,
in 1703 the viceroy, marquis of Villena, ought to state again that
public competitions were absolutely necessary and obliged those
who had been appointed lecturers without a regular competition
to undergo such a procedure. In the same occasion, Villena ordered
that every professor expounded during the competition a general
conclusion on the subject he would read (one must recall that the
lesson was divided into two parts; in the first the lecturer
dictated, in the second one he explained).\looseness-1

The Studium followed these rules until 1707, when the reign was
occupied by Austrians for twenty-seven dark years and the palace
of the Studium was occupied by the Austrian Army troops. The
lectures went back to be given again in the cloister of the
monastery of St. Domenico Maggiore, as it used to be before 1615.
The limited space made the didactics difficult but, despite the
Major Chaplain's petitions[3] to the authorities\footnote{{I.
Ascione}, \textit{L'Universit\`{a} di Napoli nei documenti del
`700 (1690-1734)} (Edizioni Scientifiche Italiane) 1997.}, the
situation did not change.

In 1735 the Reign got its independence again, the king ordered
that the seat of the Studium had to be restored, and in November 1736
the academic year was opened with an inaugural lecture given
by Giovan Battista Vico, royal lecturer of the science of rhetoric.
All the \textit{lectiones magistrales} we mentioned were given in Latin.

In 1754 something happened which changed this rule. A Tuscan mathematician,
Bartolomeno Infieri, who was living in Naples,
as administrator of the possessions
of the Medici and Corsini families, offered to the court to establish and
finance a chair with a salary of 300 ducats, gained from a bank
capital of 7500 ducats, on condition that the teaching language were
Italian. The proposal met several difficulties,
but in the end it was approved and on 5th November 1754 Antonio
Genovese obtained the chair of economic philosophy and civil
economy, which was the first chair of public economy in Europe,
giving a \textit{lectio magistralis} in front of a very large number of
people.

In the second half of the XVIII century no special changes occurred,
apart from a very slight opening to the scientific disciplines.
In 1777 the Studium was moved to bigger spaces that had been made
available after the expulsion of the Jesuits in 1767. In 1799 the Studium
was temporarily closed after the ``glorious'' Christian
army of the Cardinal Ruffo di Calabria entered Naples and
repressed the Neapolitan revolution when seven professors
were hung and eleven arrested.

The transition to a modern university was performed in 1806, when the French arrived,
by Giuseppe Bonaparte who established in the University of Naples
the classes of law, theology, medicine, natural sciences,
different chairs, and philosophy ---the latter associated with the chairs of
logic and metaphysics, of elementary mathematics, transcendental mathematics,
mechanics, experimental physics and astronomy; there was finally a
class of different chairs.

Even more modern was the innovation introduced by Gioacchino Murat,
who on the basis of an accurate analysis
carried out by a committee
of which Vincenzo Cuoco was a member, introduced, with a decree in
1811, the Faculties, among which physical and mathematical
sciences, with the chairs of synthetic mathematics, analytical
mathematics, calculus of the infinities, heuristic art or art of the
mathematical invention, mechanics, experimental physics (with a
laboratory and an assistant), zoology, botany (with a botanic
garden), vegetal physiology, natural history (with a compulsory
course of compared anatomy and with a museum kept by a professor
who was in charge of the class in natural iconography), mineralogy
(with a mineralogical cabinet and a laboratory), chemistry (with
a cabinet and an assistant in charge of the class of pharmacy) and
astronomy (with an observatory and two assistants).

In the same decree Murat reformed the competitions in the following
way:\\
-- different boards of examiners were formed
according to the disciplines' type: for instance, science was joined to
medicine (formerly there was only one board of examiners for
all the disciplines);\\
-- applicants sent the Chancellor a written
paper, in which they explained their experiences and ideas, put inside
a folder containing a sealed envelope with their
name;\\
-- the written paper was examined by a secretary:
if the paper was disapproved the envelope was burnt;\\
-- the authors of the approved papers had to
undergo an examination similar to those which were used in the past;\\
-- in the end, the members of the commission
wrote their secret mark and the result was sent to the governmental
authorities and from them to the king who signed the nomination
decree.

It was certainly kept the rule according to which the new professor
read his inaugural lesson (\textit{lectio magistralis}) in front
of the members of the faculty and invited people. This tradition
has been kept until the second world war, but it was gradually
abandoned (in Salamanca already in 1973). For instance, no one of the physicists
who won the chair after Majorana read the inaugural lesson, whereas in
the humanistic faculty some were still given in 1992 and 1993.

\vspace{-6pt}

\section*{\textbf{4. Acknowledgements.}}

The editors are grateful, for many discussions
or kind collaboration, to Franco G.Bassani, Viviano Domenici, Salvatore Esposito, 
Angela Oleandri, Emanuele Rimini, the Italian Physical Society, and particularly to 
Carmen Vasini.

\vspace{24pt}
\begin{flushright}
\begin{minipage}[b]{50mm}
{\sc Bruno Preziosi} \\
{\em Universit\`a di Napoli}\vspace{3pt}\\
{\sc Erasmo Recami}\\
{\em Universit\`a di Bergamo}
\end{minipage}
\end{flushright}

\newpage

\

\begin{center}

{\large {\bf La Lezione Inaugurale di Ettore Majorana al suo corso
di\\
Fisica Teorica}}

{(Napoli, Italia; 13 Gennaio 1938)}

\

{\large {\em {con commenti dei curatori, Bruno Preziosi ed Erasmo
Recami:}}} \ \protect\footnote{Indirizzi e-mail: recami@mi.infn.it
[ER]; preziosi@unina.it [BP]}

\end{center}

\vs{1mm}

\cent{ Bruno Preziosi, }

\vs{1mm}

\centerline{{\em Universit\`a di Napoli, Napoli, Italia;}}
 \cent{{\rm and} {\em INFM-Sezione di Napoli, Napoli, Italia.}}

\vs{1mm}

\centerline{\rm ed}

\vs{1mm}

\cent{ Erasmo Recami }

\vs{1mm}

\cent{{\em Universit\`a di Bergamo, Bergamo, Italia;}}
\cent{{\rm and} {\em INFN---Sezione di Milano, Milano, Italia.}}

\vs{8mm}

{{\bf Sommario della versione in Italiano \ --} \ In occasione
dell'anno centenario della nascita del fisico teorico Ettore
Majorana (il quale nacque a Catania il 6 agosto 1906), abbiamo
pensato di poter contribuire a sottolineare degnamente l'attuale
ricorrenza ripubblicando gli appunti preparati dal Majorana per la
lezione inaugurale del suo Corso di fisica teorica preso
l'universit\`a di Napoli, ove era stato nominato professore
ordinario per meriti eccezionali. Tale Prolusione ebbe luogo il 13
gennaio 1938. Questi appunti mostrano l'interesse dello
scienziato, non solo per le questioni generali e di fondo che
animano la ricerca scientifica, ma anche per il migliore metodo
didattico da seguire per trasmettere il sapere agli allievi (per i
]quali allievi nutriva il pi\`u profondo interesse). [Il Majorana
era tanto interessato ad insegnare quello che la sua mente
prodigiosa veniva comprendendo e scoprendo delle leggi del Creato,
che anche durante gli anni del suo presunto isolamento dal mondo e
dai colleghi,  ovvero durante gli anni accademici 1933-34, 1934-35
e 1935-36, egli present\`o domanda al fine di potere tenere dei
corsi universitari ``liberi" (e gratuiti) presso l'Istituto di
Fisica romano di Via Panisperna: cosa che gli era permessa, avendo
egli gi\`a conseguito quella che si chiamava Libera Docenza. Il
direttore, professor Orso Mario Corbino, fece approvare tali
richieste; ma non risulta che Ettore abbia tenuto quei corsi:
forse gli studenti che si rendevano conto dell'importanza degli
argomenti relativi erano troppo pochi,  in quegli anni... Ci\`o lo
si \`e saputo solo recentemente].

   Una lettura degli appunti di Majorana per la sua Prolusione pu\`o
riuscire rivelatrice, dunque,  riguardo a vari aspetti del carattere
scientifico ed umano del Nostro; avvertiamo solo che in essi ci si
riferisce alla fisica classica e alla meccanica quantistica,
trascurando in questa prima fase gli aspetti relativistici: aspetti
che verranno trattati dal Majorana solo nella seconda parte del corso,
come rivelato dagli appunti delle sue ultime sei lezioni recentemente
scoperti (le precedenti dieci lezioni, quelle allora note, furono
pubblicate per la prima volta nel 1987 presso Bibliopolis, Napoli,
insieme coi presenti appunti per la Prolusione al Corso). Il relativo
manoscritto e` stato rinvenuto da uno dei presenti curatori [ER]
nel gi\`a lontano marzo del 1973,
ma sono restate finora poco conosciute [nonostante che siano state
pubblicate dallo stesso ER nel 1982 sulle pagine della scienza del
quotidiano milanese "Corriere della Sera"].

  Ci si permetta anche di ricordare che quasi tutto il materiale
  biografico riguardante
  Ettore Majorana (fotografie incluse) \`e coperto da copyright a favore
  di Maria Majorana in solido con uno dei curatori [ER] (e con
  l'Editore
  Di Renzo [www.direnzo.it]) e non pu\`o essere ulteriormente riprodotto senza
  il consenso scritto dei detentori dei diritti.}\\

\newpage

{\bf ETTORE MAJORANA:}

{\large{\bf Gli appunti per la Lezione Inaugurale}}

\large{\rm Universit\`a di Napoli, 13 gennaio 1938}

\hyphenation{ma-gi-stra-lis bi-blio-gra-fia bi-blio-gra-fie
  ma-te-ma-ti-co ma-te-ma-ti-ci}

\

In questa prima lezione di carattere introduttivo illustreremo
brevemente gli scopi della fisica moderna e il significato
dei suoi metodi, soprattutto in quanto essi hanno di pi\`{u} inaspettato
e originale rispetto alla fisica classica.

La fisica atomica, di cui dovremo principalmente occuparci,
nonostante le sue numerose e importanti applicazioni pratiche
---e quelle di portata pi\`{u} vasta e forse rivoluzionaria che
l'avvenire potr\`{a} riservarci---, rimane anzitutto una scienza
di enorme interesse \textit{speculativo}, per la profondit\`{a}
della sua indagine che va veramente fino all'ultima radice dei
fatti naturali. Mi sia perci\`{o} consentito di accennare in primo
luogo, senza alcun riferimento a speciali categorie di fatti
sperimentali e senza l'aiuto del formalismo matematico,
ai caratteri generali della concezione della natura che \`{e} accettata
nella nuova fisica.\looseness-1
\vspace{-3pt}
\begin{center}
* * *
\end{center}
\vspace{-3pt}

La \textit{fisica classica} di Galileo e Newton all'inizio
del nostro secolo \`e interamente legata, come si sa, a quella
concezione \textit{meccanicistica} della natura che dalla fisica
\`{e} dilagata non solo nelle scienze affini, ma anche nella biologia
e perfino nelle scienze sociali, informando di s\'{e} quasi tutto
il pensiero scientifico e buona
parte di quello filosofico in tempi a noi abbastanza vicini;
bench\'{e}, a dire il vero, l'utilit\`{a}
del metodo matematico che ne costituiva la sola valida giustificazione
sia rimasta sempre circoscritta esclusivamente alla fisica.

Questa concezione della natura poggiava sostanzialmente su
due pilastri: l'esistenza oggettiva e indipendente della
materia, e il determinismo fisico. In entrambi i casi si tratta,
come vedremo, di nozioni derivate dall'esperienza comune e poi
generalizzate e rese universali e infallibili soprattutto per
il fascino irresistibile che anche sugli spiriti pi\`{u} profondi
hanno in ogni tempo esercitato le leggi esatte della fisica,
considerate veramente come il segno di un assoluto e la rivelazione
dell'essenza dell'universo: i cui segreti, come gi\`a affermava
Galileo, sono scritti in caratteri matematici.

L'\textit{oggettivit\`{a}} della materia \`{e}, come dicevo,
una nozione dell'esperienza comune, poich\'{e} questa insegna che
gli oggetti materiali hanno un'esistenza a s\'{e}, indipendente
dal fatto che essi cadano o meno sotto la nostra osservazione.
La fisica matematica classica ha aggiunto a questa constatazione
elementare la precisazione o la pretesa che di questo mondo oggettivo
\`{e} possibile una rappresentazione mentale completamente adeguata
alla sua realt\`{a}, e che questa rappresentazione mentale
pu\`{o} consistere nella conoscenza
di una serie di grandezze numeriche sufficienti a determinare
in ogni punto dello spazio e in ogni istante lo stato dell'universo
fisico.

Il \textit{determinismo} \`{e} invece solo in parte una nozione
dell'esperienza comune. Questa d\`{a} infatti al riguardo delle
indicazioni contraddittorie. Accanto a fatti che si succedono
fatalmente, come la caduta di una pietra abbandonata nel vuoto,
ve ne sono altri ---e non solo nel mondo biologico--- in cui la
successione fatale \`{e} per lo meno poco evidente. II determinismo
in quanto principio universale della scienza ha potuto perci\`{o}
essere formulato solo come generalizzazione delle leggi che reggono
la meccanica celeste. \`E ben noto che un \textit{sistema} di
punti ---quali, in rapporto alle loro enormi distanze, si possono
considerare i corpi del nostro sistema planetario--- si muove
e si modifica obbedendo alla legge di Newton.
Questa afferma che l'accelerazione di uno di questi punti si ottiene
come somma di tanti vettori quanti sono gli\ altri punti:

$$
\ddot{\overrightarrow{P_r}} \propto \Sigma_s\frac{m_s}{%
R_{rs}^2}\overrightarrow{e_{rs}},
$$

essendo $m_{s}$ la massa di un punto generico e $%
\overrightarrow{e_{rs}}$ il vettore unitario diretto da $%
\overrightarrow{P_r}$ a $\overrightarrow{P_s}$. Se in
tutto sono presenti $n$ punti, occorreranno $3n$
coordinate per fissarne la posizione e la legge di Newton stabilisce fra
queste grandezze altrettante equazioni differenziali del secondo ordine il
cui integrale generale contiene $6n$ costanti arbitrarie. Queste si possono
fissare assegnando la posizione e le componenti della velocit\`{a} di
ciascuno dei punti all'istante iniziale.
Ne segue che la configurazione futura del \textit{sistema} pu\`{o}
essere prevista con il calcolo purch\'{e} se ne conosca lo stato
iniziale cio\`{e} l'insieme delle posizioni e velocit\`{a} dei punti
che lo compongono. Tutti sanno con quale estremo rigore le
osservazioni astronomiche abbiano confermato l'esattezza della
legge di Newton; e come gli astronomi siano effettivamente in
grado di prevedere con il suo solo aiuto, e anche a grandi distanze
di tempo, il minuto preciso in cui avr\`{a} luogo un'eclisse, o
una congiunzione di pianeti o altri avvenimenti celesti.
\vspace{-3pt}
\begin{center}
* * *
\end{center}
\vspace{-3pt}

Per esporre la \textit{meccanica quantistica} nel suo stato
attuale esistono due metodi pressoch\'{e} opposti. L'uno \`{e} il
cosiddetto metodo storico: ed esso spiega in qual modo, per indicazioni
precise e quasi immediate dell'esperienza, sia sorta la
prima idea del nuovo formalismo; e come questo si sia successivamente
sviluppato in una maniera obbligata assai pi\`{u} dalla necessit\`{a}
interna che non dal tenere conto di nuovi decisivi fatti sperimentali.
L'altro metodo \`{e} quello matematico, secondo il quale il formalismo
quantistico viene presentato fin dall'inizio nella sua pi\`{u}
generale e perci\`{o} pi\`{u} chiara impostazione, e solo successivamente
se ne illustrano i criteri applicativi. Ciascuno di questi due
metodi, se usato in maniera esclusiva, presenta inconvenienti
molto gravi.

\`E un fatto che, quando sorse la meccanica quantistica, essa
incontr\`{o} per qualche tempo presso molti fisici sorpresa, scetticismo
e perfino incomprensione assoluta, e ci\`{o} soprattutto
perch\'{e} la sua consistenza logica, coerenza e sufficienza
appariva, pi\`{u} che dubbia, inafferrabile. Ci\`{o} venne anche,
bench\'{e} del tutto erroneamente, attribuito a una particolare
oscurit\`{a} di esposizione dei primi creatori della nuova meccanica,
ma la verit\`{a} \`{e} che essi erano dei fisici, e non dei
matematici,
e che per essi l'evidenza e giustificazione
della teoria consisteva soprattutto nell'immediata applicabilit\`{a}
ai fatti sperimentali che l'avevano suggerita. La formulazione
generale, chiara e rigorosa, \`{e} venuta dopo, e in parte per
opera di cervelli matematici. Se dunque noi rifacessimo semplicemente
l'esposizione della teoria secondo il modo della sua apparizione
storica, creeremmo dapprima inutilmente uno stato di disagio
o di diffidenza, che ha avuto la sua ragione d'essere ma che
oggi non \`e pi\`{u} giustificato e pu\`{o} essere risparmiato. Non
solo, ma i fisici ---che sono giunti, non senza qualche pena,
alla chiarificazione dei metodi quantistici attraverso le esperienze
mentali imposte dal loro sviluppo storico--- hanno quasi
sempre sentito a un certo momento il bisogno di una maggiore
coordinazione logica, di una pi\`{u} perfetta formulazione dei
princ\'{\i}pi, e non hanno sdegnato per questo compito l'aiuto
dei matematici.

Il secondo metodo, quello puramente matematico, presenta
inconvenienti ancora maggiori. Esso non lascia in alcun
modo intendere la genesi del formalismo e in conseguenza il posto
che la meccanica quantistica ha nella storia della scienza. Ma
soprattutto esso delude nella maniera pi\`{u} completa il desiderio
di intuirne in qualche modo il significato fisico, spesso cos\`{\i}
facilmente soddisfatto dalle teorie classiche. Le applicazioni,
poi, bench\'{e} innumerevoli, appaiono rare, staccate, perfino
modeste di fronte alla sua soverchia e incomprensibile generalit\`{a}.

Il solo mezzo di rendere meno disagevole il cammino a chi
intraprende oggi lo studio della fisica atomica, senza nulla
sacrificare della genesi storica delle idee e dello stesso linguaggio
che dominano attualmente, \`{e} quello di premettere un'esposizione
il pi\`{u} possibile ampia e chiara degli strumenti matematici
essenziali della meccanica quantistica, in modo che essi siano
gi\`a pienamente familiari quando verr\`{a} il momento di usarli
e non spaventino allora o sorprendano per la loro novit\`{a}: e
si possa cos\`{\i} procedere speditamente nella derivazione
della teoria dai dati dell'esperienza.

Questi strumenti matematici in gran parte preesistevano al
sorgere della nuova meccanica (come opera disinteressata di matematici
che non prevedevano un cos\`{\i} eccezionale campo di applicazione),
ma la meccanica quantistica li ha ``sforzati'' e ampliati per
soddisfare alle necessit\`a pratiche; cos\`{\i} essi non verranno
da noi esposti con criteri da matematici, ma da fisici. Cio\`{e}
senza preoccupazioni di un eccessivo rigore formale, che non
\`{e} sempre facile a raggiungersi e spesso del tutto impossibile.

La nostra sola ambizione sar\`{a} di esporre con tutta la chiarezza
possibile l'uso effettivo che di tali strumenti fanno i fisici
da oltre un decennio, nel quale uso ---che non ha mai condotto
a difficolt\`{a} o ambiguit\`{a}--- sta la fonte sostanziale della
loro certezza.

\newpage


\qquad
\vskip50pt

\cent{\Large{\bf COMMENTI SU ``GLI APPUNTI PER LA}}

\cent{\Large{\bf LEZIONE INAUGURALE''.}}

\vskip40pt
\setcounter{footnote}{0}

\section*{\textbf{1. Majorana: ll conferimento della cattedra
e la sua \textit{lectio magistralis}}}

 Dopo il concorso del 1926, in cui ottennero la cattedra Fermi,
Persico e Pontremoli, passarono dieci anni prima che si aprisse,
nel 1937, un nuovo concorso per la fisica teorica, richiesto
dall'universit\`{a} di Palermo per iniziativa di Emilio Segr\'{e}.
A questo nuovo concorso volle partecipare Ettore Majorana (o per
propria iniziativa o perch\'{e} invitato da amici). Per chi
conosce il carattere di Majorana, cos\`{\i} lontano da interessi
accademici, questa decisione pu\`{o} sembrare strana. Ma una
spiegazione ci \`{e} giunta nei mesi scorsi. Premettiamo il
ricordo che, dopo il rientro da Lipsia della fine del 1933, Ettore
si allontan\`{o} dal gruppo di Fermi, ma non dalla fisica, come
testimoniano[4,5] molti documenti\footnote{{E. Recami}, \textit{Il
Caso Majorana: Epistolario,
    Documenti, Testimonianze} (Mondadori, Milano) 1987, 1991; si veda
  la IV edizione ampliata (Di Renzo Editore, Roma) 2002.}.
Per di pi\`{u} De Gregorio\footnote{{A. De Gregorio} e {S.
Esposito}, in
  \textit{Sapere}, no. 3, Giugno 2006, 56; e
\textit{Teaching theoretical physics: The cases of E. Fermi and
  E. Majorana}, preprint \texttt{arXiv:phisics/0602146}.}
ha recentemente
scoperto presso l'Universit\`{a} di Roma ``La Sapienza'', che, negli
anni di isolamento, e precisamente per gli AA. AA. 1933/34, 1934/35
e 1935/36, il Majorana aveva chiesto di potere tenere presso l'Istituto
di via Panisperna dei corsi universitari ``liberi'', cosa cui
aveva diritto possedendo egli la libera docenza. Il direttore
Corbino fece approvare tali domande, ma pare che il Nostro non
tenne mai le desiderate lezioni, probabilmente per la mancanza,
allora, di studenti capaci di comprenderne la importanza.

Majorana era molto interessato, quindi, all'insegnamento di quanto
la sua mente prodigiosa andava scoprendo delle leggi della natura.
Ed \`{e} probabile che partecip\`{o} volentieri al concorso del
1937 proprio per avere finalmente degli allievi (ai quali
prest\`{o} molta attenzione, come stiamo per vedere). Come
sappiamo[4,6], su proposta della Commissione preposta al concorso,
presieduta da Fermi, il 2 novembre 1937 il ministro Bottai emette
il decreto di nomina di Ettore Majorana a professore ordinario di
fisica teorica, presso la Regia Universit\`{a} di Napoli, fuori
concorso; e alla fine del 1937 tale nomina viene partecipata dal
Ministero a Ettore, presso la sua abitazione di viale Regina
Margherita 37 in Roma, \textit{``per l'alta fama di singolare
perizia cui Ella \`{e} pervenuta nel campo degli studi riguardanti
la detta disciplina, con decorrenza dal 16 novembre 1937-XVI''.}
Majorana si reca a Napoli dopo l'Epifania (verso il 10 gennaio
1938), e il 12 scrive dalla sua sede universitaria al ministro
Bottai asserendo, tra l'altro, \textit{``\ldots\  tengo ad
affermare che dar\`{o} ogni mia energia alla scuola e alla scienza
italiane, oggi in cos\`{\i} fortunata ascesa''}\footnote{I
documenti,
  scoperti, raccolti, e per primo pubblicati,
da E. Recami, sono contenuti in bibliografia [4]. Essi (fotografie
incluse) sono coperti da copyright a favore di Recami, della
famiglia Majorana, ed, ora, dell'editore Di Renzo, ed abbisognano
del permesso scritto degli aventi diritto per la loro
riproduzione. Ovviamente ne sono escluse tutte le carte
\textit{scientifiche.}}.

Le lettere del 1938 di Ettore Majorana, aventi rilevanza anche per
le circostanze della sua scomparsa, sono contenute in bibliografia
[4]. Accenniamo brevemente solo a quelle che qui ci interessano.
Nella lettera dell'11 gennaio 1938 da Napoli, alla madre, Ettore
scrive: \textit{``Ho annunziato l'inizio del corso per
gioved\`{\i} 13 alle ore nove. Ma non \`{e} stato possibile
verificare se vi sono sovrapposizioni d'orario, cos\`{\i} che
\`{e} possibile che gli studenti non vengano e che si debba
rimandare. Ho visto il preside con cui ho concordato di evitare
ogni carattere ufficiale all'apertura del corso, e anche per
questo non vi consiglierei di venire\ldots}''. La famiglia,
invece, si present\`{o} puntuale il gioved\`{\i} 13 gennaio 1938,
alle ore nove, per assistere alla prolusione di Ettore. Alla
lezione assistettero i professori della Facolt\`{a}, fra cui
sicuramente Antonio Carrelli e Renato Caccioppoli, molto amici tra
loro; come ricorda Gilda Senatore, gli studenti non furono
invitati.

 Gli appunti per la prolusione al corso, o lezione inaugurale,
sono stati rinvenuti da uno di noi [E.R.] verso il 1972 e resi
noti per la prima volta[7] dieci anni dopo\footnote{{E. Recami},
in \textit{Corriere della Sera} (Milano), 19 Ottobre 1982. Si
vedano anche le bibliografie [4] e [8].} \footnote{{B. Preziosi}
(Curatore), \textit{Ettore Majorana -- Lezioni all'Universit\`{a}
di Napoli},
 (Bibliopolis, Napoli) 1987. Questo volume
contiene, oltre a un commento di N. Cabibbo, anche un articolo di
E. Recami contenente il gi\`{a} citato testo della lezione
inaugurale e il catalogo dei manoscritti scientifici inediti del
Nostro (ad opera di M. Baldo, R. Magnani e E. Recami); per questo
catalogo si veda anche {E. Recami}, \textit{Quaderni di Storia
della Fisica}, no. \textbf{5} (1999) 19-68, e-print
physics/9810023}:
essi sono pi\`u sopra riportati. 
In essi traspare l'interesse dello scienziato, non
solo per le questioni generali e di fondo che animano la ricerca
scientifica, ma anche per il migliore \textit{metodo didattico} da
seguire per trasmettere il sapere agli allievi (per i quali nutriva,
ripetiamo, profondo interesse).

Una lettura degli appunti di Majorana per la sua prolusione
pu\`{o} riuscire rivelatrice riguardo a vari aspetti del carattere
scientifico ed umano del Nostro; avvertiamo solo che in essi ci si
riferisce alla fisica classica e alla meccanica quantistica,
trascurando in questa prima fase gli aspetti relativistici:
aspetti che verranno trattati dal Majorana solo nella seconda
parte del corso, come rivelato dagli appunti delle sue ultime sei
lezioni recentemente scoperti. Majorana era particolarmente
sedotto dalla descrizione anti-meccanicistica e ``probabilistica''
della meccanica quantistica, tanto da trattarla ampiamente[9]
anche nel suo articolo\footnote{{E. Majorana}, ``Il valore delle
leggi statistiche nella fisica e nelle scienze sociali'',
\textit{Scientia} \textbf{36} (1942) 58-66. }, pubblicato postumo,
nel 1942, da Giovannino Gentile. Gi\`a il 27 luglio 1934, da
Monteporzio Catone (RM), Majorana aveva scritto allo stesso
Gentile: \textit{``Credo che il maggior merito del libro} [10]
\textit{di Jeans}\footnote{J. Jeans, \textit{I Nuovi Orizzonti
della Scienza} (Sansoni, Firenze) 1934, traduzione italiana a cura
di G. Gentile jr.} \textit{sia quello di anticipare le reazioni
psicologiche che il recente sviluppo della fisica dovr\`{a}
fatalmente produrre quando sar\`{a} generalmente compreso che la
scienza ha cessato di essere una giustificazione per il volgare
materialismo\ldots''}.

Poich\'{e} il mito ha associato la scomparsa di Ettore con
timori circa la possibile costruzione della bomba atomica, osserviamo
subito che, fin dagli inizi della sua lezione inaugurale, Ettore
dichiara esplicitamente:\textit{``\ldots\ La fisica atomica, di cui dovremo
principalmente occuparci, nonostante le sue numerose e importanti
applicazioni pratiche ---e quelle di portata pi\`{u} vasta e forse
rivoluzionaria che l'avvenire potr\`{a} riservarci---,
rimane anzitutto 
una scienza di enorme interesse speculativo, per la profondit\`{a}
della sua indagine che va veramente fino all'ultima radice dei
fatti naturali\ldots}''. Il periodare di Majorana lascia intendere
che, anche di fronte alle applicazioni forse ``rivoluzionarie''
alle quali la fisica atomica e nucleare avrebbero potuto portare,
il loro interesse (in particolare per lui) \`{e} essenzialmente
quello speculativo.

\vspace{-6pt}

\section*{\textbf{2. Le lezioni del suo corso di fisica teorica}}

Come testimoniato da Gilda Senatore e Sebastiano Sciuti, gli
alunni del corso, che inizi\`{o} il 15 gennaio, furono, oltre a
loro due, Nella Altieri, Laura Mercogliano, Nada Minghetti e
Savino Coronato, allievo di Caccioppoli, che dopo l'ultima lezione
non frequent\`{o} pi\`{u} l'Istituto Fisico e che si laure\`{o} in
Matematica lo stesso anno. A loro testimonianza nessun altro
partecip\`{o}, salvo, molto sporadicamente, Mario Cutolo, gi\`{a}
laureato in fisica.

Ai propri studenti il Majorana dedicava la pi\`{u} grande attenzione;
e ne era soddisfatto. Invero, il 2 marzo 1938, nella sua ultima
lettera all'amico e collega Giovanni Gentile jr, scrive:
``\textit{\ldots\ Sono contento degli studenti, alcuni dei quali
sembrano risoluti a prendere la fisica sul serio\ldots}''.
Quando prendeva in mano il gesso,
la sua timidezza scompariva ed Ettore, come \`{e} facile immaginare,
si trasfigurava, mentre dalla sua mano uscivano con facilit\`{a}
intere, eleganti lavagne di simboli fisici e matematici. Ci\`{o}
\`{e} stato ricordato da Gilda Senatore di fronte alla telecamera di Bruno Russo, e pi\`{u}
di recente, in occasione del 60mo anniversario dalla sua scomparsa,
in un convegno organizzato dall'Universit\`a Federico II di Napoli.

 L'intera serie degli appunti autografi di lezione redatti con
ogni cura da Majorana, a beneficio dei propri allievi (e forse
Ettore stava pensando di scrivere un libro per studenti, cos\`{\i}
come aveva pensato ad un libro nello stendere i suoi
originalissimi appunti di studio[11],
i~\textit{Volumetti}\footnote{{S. Esposito}, {E. Majorana} jr, {A.
van der Merwe} e {E. Recami}, \textit{Ettore Majorana - Notes on
Theoretical Physics,} (Kluwer Academic Press, Dordrecht, Boston e
New York) 2003. (Edizione nella lingua originale italiana:
{Majorana E.}, \textit{Appunti inediti di fisica teorica}, a cura
di S. Esposito e E. Recami (Zanichelli, Bologna) 2006).}) fu
consegnata alla prediletta studentessa Gilda Senatore insieme con
altri scritti, che non sono stati ritrovati, il giorno prima di
partire per Palermo. Come queste carte arrivarono, tra la fine del
'38 ed i primi del '39, a Carrelli e in che occasione Carrelli le
trasmise ad Amaldi, prive di sei lezioni riguardanti
l'elettrodinamica e la relativit\`{a} speciale[12], \`{e}
descritto in~\footnote{\textit{L'eredit\`{a} di Fermi e Majorana
ed altri temi} (Bibliopolis, Napoli) 2006.} e in una lettera di
Preziosi a ``Le Scienze'' (settembre 2002). \`E interessante
notare che nel 1939-40 Carrelli tenne un corso di relativit\`a
speciale, con le relative dispense pubblicate dal GUF nel 1940. Le
dieci lezioni trasmesse ad Amaldi, e da lui depositate alla Domus
Galilaeana, furono pubblicate anastaticamente in [8].
Recentemente[13], S. Esposito\footnote{S. Esposito, \textit{Nuovo
    Saggiatore}, \textbf{21} No. 1-2 (2005) 21-41.}
ed Antonino Drago hanno rinvenuto, fra le carte lasciate alla
famiglia da Eugenio Moreno, uno studente di Matematica che si
laure\`o con Caccioppoli nel 1941, la trascrizione, di pugno del
Moreno, di tutti gli appunti manoscritti da Majorana, incluse la
parte di relativit\`{a} che non c'\`{e} tra i documenti depositati
nella Domus. Tali appunti completi[14] sono in~\footnote{{S.
Esposito } (Curatore), \textit{Ettore Majorana - Lezioni di Fisica
Teorica} (Bibliopolis, Napoli) 2006.}.


\vspace{-6pt}

\section*{\textbf{3. Le procedure della chiamata e della \textit{lectio magistralis} a
Napoli}}

Il 5 giugno del 1224 Federico II, re di Germania ed imperatore
dei Romani, inviava da Siracusa a tutte le autorit\`{a} del Regno
una circolare (\textit{generales licterae}) che esordiva con:\looseness-1

\textit{``Col favore di Dio, per il quale viviamo e regniamo, al quale
riferiamo quanto di bene facciamo, desideriamo che, mediante
una fonte di scienza ed un semenzaio di dottrina, nel Regno nostro
molti diventino savi ed accorti, i quali, resi abili dallo studio
e dalla meditazione del diritto, servano a Dio, al quale tutte
le cose servono, e piacciano a noi per il culto della giustizia,
ai cui precetti ordiniamo a tutti di obbedire. Abbiamo perci\`{o}
disposto che, nell'amenissima citt\`{a} di Napoli, s'insegnino
le arti e si coltivino gli studi di ogni professione, affinch\'{e}
i digiuni ed affamati di dottrina trovino dentro il Regno stesso
di che soddisfare le loro brame, e non sieno costretti, per procurare
d'istruirsi, a imprendere lunghi viaggi, e mendicare in terre
straniere.''}

Nella stessa circolare[1] annunziava che \textit{``uno dei maestri
da lui scelti sarebbe stato il dottissimo Roffredo di
Benevento''}, e stabiliva che \textit{``si far\`{a} prestito agli
scolari}\ldots''; come riferito da Torraca\footnote{Stamperia di
Giovanni de Simone, Napoli MDCCLIV.}, le lezioni allo ``Studio''
sarebbero cominciate nell'ottobre del 1224.

La nomina dei professori e la durata della stessa erano quindi
prerogativa dell'imperatore. Tale procedura fu mantenuta da Corrado
e Manfredi e, dopo una breve interruzione succeduta alla battaglia di
Benevento (1266), dagli Angioini (1266-1442) e successivamente
dagli Aragonesi (1442-1503).
\`E nel 1503 che ha inizio il periodo spagnolo con Ferdinando
il Cattolico. All'inizio lo Studio rimase chiuso per qualche
anno, ma fu riaperto il giorno di S. Luca (18 Ottobre) del 1507.
Va notato che era in tale giorno che l'Universit\`{a} di Salamanca,
fondata sei anni prima dello Studio, usava riaprire i battenti;
peraltro, come vedremo pi\`{u} avanti, lo Studio di Salamanca fu
sempre un riferimento per Napoli.\looseness-1

La persona che port\`{o} un significativo contributo fu senz'altro
il vicer\`{e} D.~Pietro Fernandez de Castro, conte di Lemos. Questi,
infatti, non solo fece edificare un grande edificio fuori della
porta di Costantinopoli, ora sede del Museo Archeologico, che
fu occupato dallo Studio dal 1615 fino all'inizio del secolo
XVIII, ma realizz\`{o}, fra il 1614 e il 1616, una profonda riforma
nella procedura per il reclutamento dei professori. Tale riforma
copia sostanzialmente la regola sancita per l'Universit\`{a} di
Salamanca nel 1561, in cui si stabilisce che il reclutamento
avviene con una procedura di concorso bandito dal Governo e dopo
un pubblico esame davanti ad una commissione composta da professori
e lettori, anche di collegi religiosi. A Salamanca, dopo il pubblico
esame gli aspiranti attendevano in una cappella la chiamata del
vincitore e l'invito ad unirsi al consesso dei professori.\looseness-1

Come riportato da Giangiuseppe Origlia[2] nella sua
\textit{Istoria dello Studio di Napoli}\footnote{{F. Torraca},
\textit{Storia della Universit\`{a} di Napoli} (Riccardo Ricciardi
Editore, Napoli) MCMXXIV.}, il pubblico esame consisteva
\textit{``in porre al concorrente l'obbligo di pubblicamente
sporre a viva voce e per lo continuo spazio di un'ora e senza
l'aiuto de' scritti \ldots, quei punti della materia, \ldots, li
quali 24 ore prima''} gli erano stati assegnati \textit{``dal
Prefetto in} \textit{presenza de' testimoni.''}

A dimostrazione dell'interesse con cui queste lezioni erano seguite,
Origlia precisa che il pubblico presente era costituito dai
\textit{``lettori e tutti quelli che avevano la facolt\`{a} di dare il
suffragio alla} \textit{Cattedra, ch'era da conferirsi, non che d'uno
stuolo infinito de' scolari, o d'altri, che desideravano in simili
giostre essere presenti''.}

Il Lemos stabil\`{\i} anche le regole per l'apertura dell'Anno
Accademico (la prima il 14 giugno 1615). Secondo una testimonianza
di un contemporaneo, la cerimonia inizi\`{o}
con un corteo in cui
\textit{``andavano prima i legisti con
  mozzetta{\renewcommand{\thefootnote}{*}\footnote{Corta mantellina.}}
  di drappo verde e cappello
con fiocco di seta verde, quindi i medici} \textit{con mozzetta di
drappo torchino e cappello con fiocco dello stesso colore, quindi
i teologi con mozzetta di drappo bianco e cappello dello stesso
colore.''}
\setcounter{footnote}{2}

Una volta giunti al palazzo dello Studio, ebbe luogo la cerimonia
con una \textit{lectio magistralis} letta da
Gio. Lorenzo di Rogiero.

Le acconciature suscitarono dei risolini tra parte del popolo;
ci\`{o} nonostante furono mantenute per le successive analoghe
circostanze.

In ref.[2], Nino Cortese descrive le occasioni in cui veniva letta
una lezione con carattere ufficiale nel seguente modo:
\textit{``L'anno scolastico si apriva solennemente con un'orazione
di uno dei lettori; inoltre si continu\`{o} un uso gi\`{a} in voga
nel secolo precedente, pel quale questi ultimi, all'atto di
prendere possesso della cattedra, pronunciavano una vera e propria
prolusione.''}

Questa prolusione inaugurale era una usanza seguita anche a Salamanca
e divenne una tradizione che in Spagna trov\`{o} una formalizzazione
in un Decreto Real del 20 agosto 1859, ma a Napoli rimase interna
all'Ateneo.

 Tornando alla procedura di reclutamento, va detto che non sempre
avvenne per concorso tant'\`{e} che il vicer\`{e}
marchese di Villena dovette, nel 1703, ribadire che i concorsi
erano assolutamente necessari ed obblig\`{o} coloro che erano stati
nominati lettori senza aver sostenuto un concorso a sottomersi
a tale procedura. Nella stessa prammatica, il Villena ordina
che ogni cattedratico esponga durante il corso una conclusione
generale della materia che legge (si tenga presente che la lezione
era divisa in due parti; nella prima il lettore dettava, nella
seconda spiegava).
\vspace{12pt}

Lo Studio segu\`{\i} queste regole sino al 1707, quando il regno
venne occupato dagli Austriaci per ventisette oscuri anni e il
Palazzo dello Studio fu occupato dalle truppe austriache. Le
lezioni tornarono a tenersi nel chiostro del convento di S.
Domenico Maggiore, come usava prima del 1615. La ristrettezza
dello spazio rese difficile lo svolgimento della didattica, ma la
situazione non cambi\`{o}, nonostante le suppliche[3] del
Cappellano Maggiore alle Autorit\`{a}\footnote{{I. Ascione},
\textit{L'Universit\`{a} di Napoli nei documenti del `700
(1690-1734)} (Edizioni Scientifiche Italiane) 1997.}.

Nel 1735 il Regno riacquist\`{o} l'indipendenza, il Re ordin\`{o}
di restaurare la sede dello Studio ed il 4 Novembre 1736 fu inaugurato
l'anno accademico con una prolusione di Giovan Battista Vico,
lettore regio della scienza della Retorica.
Tutte le \textit{lectiones magistrales} di cui si \`{e} detto erano
rigorosamento in latino.

Nel 1754 ci fu un evento che dette una svolta a questa regola.
Un matematico toscano, Bartolomeo Infieri, vivente a Napoli,
amministratore di beni dei Medici e dei Corsini, propose alla
Corte di istituire una cattedra finanziata con una rendita di 300 ducati,
frutto di un capitale in banca di 7500 ducati, a condizione che
l'insegnamento fosse impartito in lingua italiana. Questa proposta
incontr\`{o} varie difficolt\`{a}, ma alla fine fu approvata e il
5 novembre 1754 Antonio Genovese pot\`{e} salire sulla cattedra
di filosofia economica e di economia civile, che fu la prima
cattedra di economia pubblica in Europa, tenendo una \textit{lectio magistralis}
alla presenza di un pubblico straordinariamente folto.

Nella seconda met\`{a} del '700 non ci furono particolari novit\`{a},
salvo una leggerissima apertura alle discipline scientifiche,
il trasferimento dello Studio nel 1777 nei notevoli spazi resisi
liberi dopo la cacciata dei gesuiti del 1767, e la temporanea
chiusura dello Studio nel 1799 dopo l'ingresso della \textit{gloriosa
armata cristiana} del cardinal Ruffo di Calabria che represse
la Rivoluzione Napoletana, con sette professori afforcati ed
undici arrestati.\looseness-1

Il passaggio ad una Universit\`{a} moderna si ebbe con la venuta
dei francesi nel 1806 al seguito di Giuseppe Bonaparte, che
ripart\`{\i} l'Universit\`{a} degli Studi di Napoli nelle classi
di diritto, teologia, medicina, scienze naturali, e filosofia:
quest'ultima associata alle cattedre di logica e metafisica,
matematica semplice, matematica trascendentale, meccanica, fisica
sperimentale e astronomia; c'era infine una classe di cattedre
diverse. Ancora pi\`{u} moderna l'innovazione introdotta da
Gioacchino Murat, che, sulla base di una accurata analisi condotta
da una commissione di cui fu relatore Vincenzo Cuoco, con decreto
del 1811, introdusse le Facolt\`{a}, fra cui quella di scienze
fisiche e matematiche, con le cattedre di matematica sintetica,
matematica analitica, calcolo degl'infiniti, arte euristica o
dell'invenzione matematica, meccanica, fisica sperimentale (con un
gabinetto di macchine e un aggiunto), zoologia, botanica (con un
giardino botanico), fisiologia vegetale, storia naturale (con
l'obbligo del corso di anatomia comparata e con un museo curato da
un professore cui era affidato il corso di iconografia naturale),
mineralogia (con un gabinetto mineralogico ed un laboratorio),
chimica (con un gabinetto ed un aggiunto cui era affidato il corso
di farmacia) e di astronomia (con un osservatorio e due aggiunti).

Con lo stesso decreto Murat riform\`{o} i concorsi sulla base delle seguenti regole:\\
-- le commissioni d'esame non vedevano pi\`{u} insieme tutti i professori,
ma questi venivano accorpati (ad esempio, scienze era accorpata a medicina);\\
-- i candidati inviavano al Cancelliere uno scritto con l'esposizione
della propria esperienza e delle proprie idee in un plico contenente una busta al cui interno
c'era il proprio nome;\\
-- questo scritto veniva esaminato da un segretario; se lo scritto
era disapprovato, la busta veniva bruciata;\\
-- gli autori degli scritti approvati venivano sottoposti ad un
esame analogo a quelli in uso nel passato;\\
-- al termine i commissari esprimevano il loro voto segreto ed
il risultato veniva trasmesso alle autorit\`{a} di governo e
quindi al Re che firmava il decreto di nomina.\\
Certamente era mantenuta la regola secondo cui il nuovo professore
pronunciava la sua lezione inaugurale, detta anche \textit{lectio magistralis},
alla presenza della facolt\`{a} e di invitati.

Quest'ultima tradizione \`{e} sicuramente durata fino
alla seconda guerra mondiale, ma \`{e} stata man mano abbandonata
(a Salamanca gi\`a nel 1973). Ad esempio, nessuno dei fisici che sono
andati in cattedra dopo Majorana l'ha tenuta, mentre risulta
che, nella nostra facolt\`{a} di lettere, ne siano state ancora
tenute nel 1992 e nel 1993.

\vspace{-6pt}

\section*{\textbf{4. Ringraziamenti.}}

Gli ``editors" sono grati, per varie discussioni o gentile
collaborazione, a Franco G.Bassani, Viviano Domenici, Salvatore Esposito, Algela
Oleandri, Emanuele Rimini, la Societ\`a Italiana di Fisica, e in particolare a 
Carmen Vasini.

\vspace{24pt}
\begin{flushright}
\begin{minipage}[b]{50mm}
{\sc Bruno Preziosi} \\
{\em Universit\`a di Napoli}\vspace{3pt}\\
{\sc Erasmo Recami}\\
{\em Universit\`a di Bergamo}
\end{minipage}
\end{flushright}

\newpage
\section*{\textbf{5. A short Bibliography.}}

[1] ``Stamperia" of Giovanni de Simone, Neaples, Italy, MDCCLIV.

[2] \textit{Storia della Universit\`{a} di Napoli} (Riccardo
Ricciardi Editore; Neaples, Italy, MCMXXIV).

[3] Ascione I., \textit{L'Universit\`{a} di Napoli nei documenti
del `700 (1690-1734)} (Edizioni Scientifiche Italiane; 1997).

[4] Recami E., \textit{Il Caso Majorana: Epistolario, Documenti,
Testimonianze} (first editions, 1987 and 1991, by Mondadori
publisher, Milan, Italy).  See the enlarged IV edition (2002) by
Di Renzo Editore, Rome, Italy [\texttt{www.direnzo.it}].

[5] De Gregorio A. and Esposito S., in \textit{Sapere} (in press);
and \textit{Teaching theoretical physics: The cases of E. Fermi
and
  E. Majorana}, submitted for pub.

[6] Note in Italian: I documenti, scoperti, raccolti, e per primo
pubblicati, da E. Recami, sono contenuti in bibliografia [4]. Essi
(fotografie incluse) sono coperti da copyright a favore di Recami,
Maria Majorana, ed, ora, dell'editore Di Renzo, ed abbisognano del
permesso scritto degli aventi diritto per la loro riproduzione.
Ovviamente ne sono escluse tutte le carte \textit{scientifiche.} \
\ \ [Note in English: The documents, discovered collected and
first published by E.Recami, are contained in ref.[4]. They
(photos included) are proteceted by copyright in favour of Recami,
Maria Majorana, and, now, of the publisher Di Renzo; and for their
reproduction needs a written permission from the right holders. Of
course, all purely \textit{scientific} papers are of free use.]

[7] Recami E., in \textit{Corriere della Sera} (Milan, Italy), 19
October 1982.  See also refs. [4] and [8].

[8] Preziosi B. (editor,) \textit{Ettore Majorana - Lezioni
all'Universit\`{a} di Napoli}
 (Bibliopolis, Napoli) 1987. \ \ Oss. in Italian: Questo volume
contiene, oltre a un commento di N. Cabibbo, anche un articolo di
Recami comprendente il gi\`{a} citato testo della lezione
inaugurale e il catalogo dei manoscritti scientifici inediti del
Nostro (ad opera di Baldo M., Mignani R. e Recami E.); per questo
catalogo si veda anche Recami E., \textit{Quaderni di Storia della
Fisica}, no. \textbf{5} (1999), 19-68, e-print physics/9810023. \
\ [Obs. in English: That volume contains, besides a comment by
N.Cabibbo, also an article by Recami which includes the mentioned
text of the Inaugural Lecture, as well as a catalog (by Baldo,
Mignani and Recami) of the scientific manuscripts left unpublished
by E.Majorana; for such a catalog, see also Recami E.,
\textit{Quaderni di Storia della Fisica}, no. \textbf{5} (1999),
19-68, e-print physics/9810023.]

 [9] Majorana E., ``Il valore
delle leggi statistiche nella fisica e nelle scienze sociali'',
\textit{Scientia} \textbf{36} (1942) 58-66. Two English versions
exist of the present semi-popularization article.

[10] Jeans J., \textit{I Nuovi Orizzonti della Scienza} (Sansoni;
Firenze, 1934), translated into Italian by G. Gentile jr.

[11] Esposito S., Majorana E. jr, van der Merwe A. and Recami E.,
\textit{Ettore Majorana - Notes on Theoretical Physics,} (Kluwer
Academic Press, Dordrecht, Boston and New York) 2003. [Edizione
nella lingua originale italiana: \textit{Ettore Majorana - Appunti
di Fisica Teorica}, a cura di Esposito S. e Recami E. (Zanichelli,
Bologna) 2006)].

[12] \textit{L'eredit\`{a} di Fermi e Majorana ed altri temi} ,
(Bibliopolis; Neaples, Italy, 2006).

[13] Esposito S.,  \textit{Nuovo Saggiatore}, \textbf{21} (2005)
21-41.

[14] Esposito S. (editor), \textit{Ettore Majorana - Lezioni di
Fisica Teorica}, (Bibliopolis; Neaples, Italy, 2006).

\end{document}